\documentclass[12pt,a4paper]{article}
\usepackage{graphicx}
\usepackage{times}
\title{\bf Investigating the properties of Galactic Luminous Blue Variables via IR observations}

\author{J. S. Clark$^1$, A. Arkharov$^2$, V. Larionov$^3$, B. Ritchie$^1$, P. Crowther$^4$, F. Najarro$^5$\\
\vspace{1cm}\\
\normalsize $^1$ Dept. Physics \& Astronomy, The Open University, Milton Keynes, UK\\ 
\normalsize $^2$ Pulkovo Astronomical Observatory, St Petersburg, Russia\\
\normalsize $^3$ Astronomical Institute of St. Petersburg University, St Petersburg, Russia\\
\normalsize $^4$ Dept. Physics \& Astronomy, University of Sheffield, Sheffield, UK \\ 
\normalsize $^5$ Dept.  de Astrof\'{i}sica, CAB, INTA-CSIC, Torrej\'{o}n de Ardoz,  Spain
}
\date{\mbox{}}
\begin{document}
\maketitle
\pagestyle{empty}
%
% WE REDEFINE THE plain LaTeX PAGESTYLE !!! 
% THIS PAGESTYLE WILL BE USED FOR THE FIRST PAGE ONLY !
%
\def\bull{\vrule height .9ex width .8ex depth -.1ex}
\makeatletter
\def\ps@plain{\let\@mkboth\gobbletwo
\def\@oddhead{}\def\@oddfoot{\hfil\tiny\bull\quad
``The multi-wavelength view of hot, massive stars''; 39$^{\rm th}$ Li\`ege Int.\ Astroph.\ Coll., 12-16 July 2010 \quad\bull}%
\def\@evenhead{}\let\@evenfoot\@oddfoot}
\makeatother
%
% AND DEFINE OUR MACROS FOR THE REFERENCE LIST
% I.E \beginrefer \refer and \endrefer
%
\def\beginrefer{\section*{References}%
\begin{quotation}\mbox{}\par}
\def\refer#1\par{{\setlength{\parindent}{-\leftmargin}\indent#1\par}}
\def\endrefer{\end{quotation}}
%
% BEGIN THE ABSTRACT CHAPTER WITH \noindent\small, ENCLOSE IT IN A GROUP
% AND BOLDFACE THE TITLE.
%
{\noindent\small{\bf Abstract:} 
Recent IR surveys of the Galactic plane have revealed a large number of candidate Luminous Blue 
Variables. In order to verify these classifications we have been undertaking a long term 
spectroscopic and photometric monitoring campaign supplemented with tailored non-LTE model atmosphere analysis. Here we present a brief overview of
selected aspects of this program, highlighting the prospects for identification,  classification and quantitative analysis of  LBVs in the near-IR spectral window.}

%
% NOW COMES THE MAIN BODY OF THE ARTICLE
%
\section{Introduction}

Luminous Blue Variables (LBVs) are massive post-Main Sequence stars that are experiencing  a  highly unstable phase of 
evolution that is characterised by dramatic photometric and spectroscopic variability and heavy mass loss. They have been the subject of much recent interest given the twin 
possibilities that their high mass-loss  rates -- particularly during transient outbursts -- may be essential   
 for the formation of H-depleted Wolf Rayet stars (e.g. Smith \& Owocki 2006) and
 that they may be the immediate precursors of a subset of highly luminous Type II supernovae (e.g. Gal-Yam \& Leonard 2009). 

Historically, their  rarity (e.g.  Clark et al. 2005) has meant  
that their properties --  particularly regarding their characteristic outbursts and eruptions (duration, duty cycle, associated mass loss rate and underlying physical 
cause) -- have remained poorly understood. However, recent narrow- and broad-band infra-red surveys of the Galactic Plane  have
 revealed a large number of new LBVs candidates (Clark et al., 2003, Gvaramadze et al. 2010, Mauerhan et al. 2010, Wachter et al. 2010) and it is hoped that studies of an 
expanded 
sample size will help elucidate the nature of the LBV phenomenon and its role in massive stellar evolution. However, given their location in the Galactic plane, observations of these stars must be 
undertaken in the (near)-IR due to significant line of sight extinction. In this contribution
 we preview the results of a long term spectroscopic and photometric campaign of recently 
identified candidate LBVs,  supplemented with tailored model atmosphere analysis
utilising the CMFGEN code (Hillier \& Miller 1998); a  full description of this program will 
be presented in Clark et al. (in prep.).

\section{Data Reduction and Analysis}
Since 2001, near-IR JHK broadband photometric observations of our targets have been  obtained 
with  the AZT-24 1.1m telescope  in Campo Imperatore (Italy). Contemporaneous spectroscopy has
been obtained from a number of facilities  including the AZT-24 1.1m, UKIRT, the Mayall 4m and 
the VLT, while we have also made use of published spectroscopy and photometry. A full description
of data collection and reduction will be presented in Clark et al. (in prep.). 

Due to the numerous potential sources of  near-IR variability -- such as continuum emission from the stellar wind,  emission/extinction due to circumstellar 
dust and 
changes in 
stellar temperature and bolometric luminosity -- it is impossible to constrain the behaviour of 
LBVs from photometric observations alone. Consequently, where possible, we have undertaken 
quantitative modeling of the combined datasets; a description of the methodology employed is found in Clark et al. (2009). 

\section{Selected Preliminary Results}

In Fig. 1 we present sample lightcurves for two recently identified LBVs; AFGL 2298 and G24.73+0.69 (Clark et al. 2005). Both are clearly variable over $\geq$decades, with 
$\Delta$JHK$\geq$1.5~mag. As 
such, the timescales and magntitudes of variability are entirely comparable to the results of optical monitoring although, as mentioned above, 
contemporaneous spectroscopy is required to interpret these data. In the case of AFGL 2298, analysis of such a dataset revealed that its bolometric luminosity 
 varied by more than a factor  of two over the course of the observations (Clark et al. 2009). This behaviour  was  driven by significant changes in the stellar radius, which were
 accompanied by relatively small changes in temperature (Fig. 2). Recent analysis by Groh et al. (2009) demonstrated that AG Car also varied in bolometric 
luminosity over the course of its photometric  excursions, with a reduction in luminosity as the star expanded and cooled  due to the energy required 
to support the extended outer
 layers of the star against gravity. However, unlike AG Car, the maximum 
luminosity of AFGL 2298 occured when its radius was also at a maximum rather than at a minimum; the `pulsations' of both stars therefore appearing  to be of a different 
character, with those of AFGL2298 
being more reminiscent of a (weak) `Giant eruption' rather than a canonical `S Dor' excursion.

A similar dataset also exists for G24.73+0.69 and the results of a comparable quantitative analysis will be presented in a future work, although preliminary 
comparison of spectra obtained in the transition from photometric minimum (Clark et al. 2003) to maximum (Fig.2) suggests a cooling of the star in a manner analogous to AG 
Car.

\subsection {A near-IR spectral classification scheme}

Building on this approach, the identification and subsequent spectral follow up of numerous new LBV candidates  enables us to define a classification scheme
for LBVs in the near-IR as well as investigating the parameter space they occupy and their placement  in an evolutionary scheme. We present a montage of K-band spectra of 
(candidate) LBVs/WN9-11h stars   suitable for classification in Fig. 2, along with sample  WN8 and Yellow Hypergiants (YHGs) spectra and the results of non-LTE model 
atmosphere analysis where available. 

We highlight the diverse spectral morphology of (c)LBVs, as might be expected given that we sample stars with temperatures ranging from $\sim$8-20kK. Nevertheless, such stars 
are distinct from the `normal'  Blue Supergiants that also span this temperature range (Clark et al. in prep., Hanson et al. 1996) but which, as a result of their lower 
wind densities,  lack the prominent emission lines of H\,{\sc i}, He\,{\sc i} and low excitation metals that characterise the spectra of (c)LBVs.

While unfortunately no luminosity dependent features are present in this wavelength range, the presence and line profiles  of the various species do allow a 
gross, qualitative determination of stellar temperature, with the coolest (c)LBVs  (T$<$10kK) demonstrating Na\,{\sc i} doublet emission and critically  
lacking  He\,{\sc i}
2.112$\mu$m emission or absorption. At higher temperatures He\,{\sc i} 2.112$\mu$m  is  intially observed in absorption, before being driven into emission along with He\,{\sc i} 2.058$\mu$m and 
low excitation species such as Mg\,{\sc ii} and Fe\,{\sc ii}. At still higher temperatures these lines disappear, leaving a simple emission line spectrum dominated by Br$\gamma$ 
and various He\,{\sc i} lines. This process also seems to be accompanied by the development of a pronounced P Cygni profile in the He\,{\sc i } 2.058$\mu$m line.
Finally, we note that the He\,{\sc ii} 2.189$\mu$m line appears in emission for the WN8h-9h stars but is absent for the 
cooler early B supergiants such as P Cygni. It is observed to be weakly in emission in WN9h stars and to show 
a range of strengths in the WN8 stars due to the large temperature range spanned by this subtype (e.g. WR123 \& 124; 
Crowther \& Smith 1996, Crowther et al. 1999), hence it {\em may} also distinguish between these subtypes 
(e.g. Crowther et al. 2006). In this respect we note that LHO158, listed as WN8h by Liermann et al. (2009),
 could formally be classified as either WN8-9h.

However, we caution  that in the region of parameter space sampled by (c)LBVs and the closely related WN9-11h stars, the K band spectra of such stars can show a degeneracy 
whereby multiple combinations of stellar temperature, mass-loss rate and H/He ratio may result in similar spectral morphologies (e.g. Hillier et al. 1998). Indeed, this 
problem may be appreciated by noting the similarities between the spectra of P Cygni and HDE316285 in Fig. 2 despite the significant difference in temperature between the two 
stars.  We therefore emphasise that tailored, quantitative analysis of the spectra of  individual stars over as broad a wavelength range as possible (due to the comparative lack of 
diagnostics in the K band)  is required for the extraction of their physical properties.   The Pistol star is a
case in point, with Najarro et al. (2009) finding a downwards revision in luminosity by  a factor of $>$2.5 over previous estimates following such modeling.
 Unfortunately, this requires both high S/N and, critically, spectral  resolution,  given the  low terminal wind velocities -- v$\leq$500kms$^{-1}$ and typically 
$\leq$200kms$^{-1}$ -- of such stars.

Nevertheless, preliminary results from such analyses  show an encouraging continuity in physical properties -- and indeed spectral morphologies -- with the WN8 stars, with 
a steady 
march to  higher temperatures and wind  velocities at relatively constant mass-loss rates. Such a connection has already been suggested by Langer et al. (1994) and latterly 
by Martins et al. (2007) based on the evolved population of the Galactic 
Centre cluster; indeed, analysis of the properties of the (evolved) stellar populations  found within young massive clusters is a powerful tool in constraining the passage of stars from the Main Sequence through the 
`transitional' zoo (e.g. Clark et al. 2010). At the other extreme there is a striking similarity between the YHG IRC+10 420 -- which is currently evolving to higher 
temperatures -- and LBVs  in a 
cool phase such as G24.73+0.69 and G0.120-0.048. We do not claim that all LBVs evolve directly from YHGs, however; while this might be possible for low luminosity (and mass)
 stars such as G24.73+0.69, the lack of high luminosity cool super/hyper-giants  clearly indicate that stars as luminous as G0.120-0.048 could not have evolved via such a pathway. 

\section{Concluding remarks and future prospects}
While various lines of evidence suggest an important role for LBVs in the late evolutionary stages of massive stars -- and by extention their death in SNe and the nature 
of the
post-SNe relativistic remnants -- the properties of this phase are still poorly understood, in large part due to the rarity of 
such stars. However, the recent identification of large numbers of new candidates within the Galactic plane and the viability of studying them via concerted spectroscopic 
and 
photometric monitoring supplemented with tailored non-LTE model atmosphere analysis will allow these issues to be directly addressed. Indeed, such work will greatly benefit from
near-IR surveys such as VISTA/VVV  and the advent of  1-2m class robotic facilities such as the Faulkes Telescopes. Likewise, the availability  of multiplexing spectrographs 
and transient surveys such as PanSTARRS will  permit similar studies in external galaxies over a range of metalicities. When combined with 
radiative transfer  modeling of the spatially resolved gaseous \& dusty ejecta associated with large numbers of galactic  LBVs and which encodes their past mass loss 
histories, 
these programs have the potential to advance studies of this transient and violent phase of stellar evolution over the coming years.

\begin{figure}[h]
\begin{minipage}{8cm}
\centering
\includegraphics[width=6cm,angle=270]{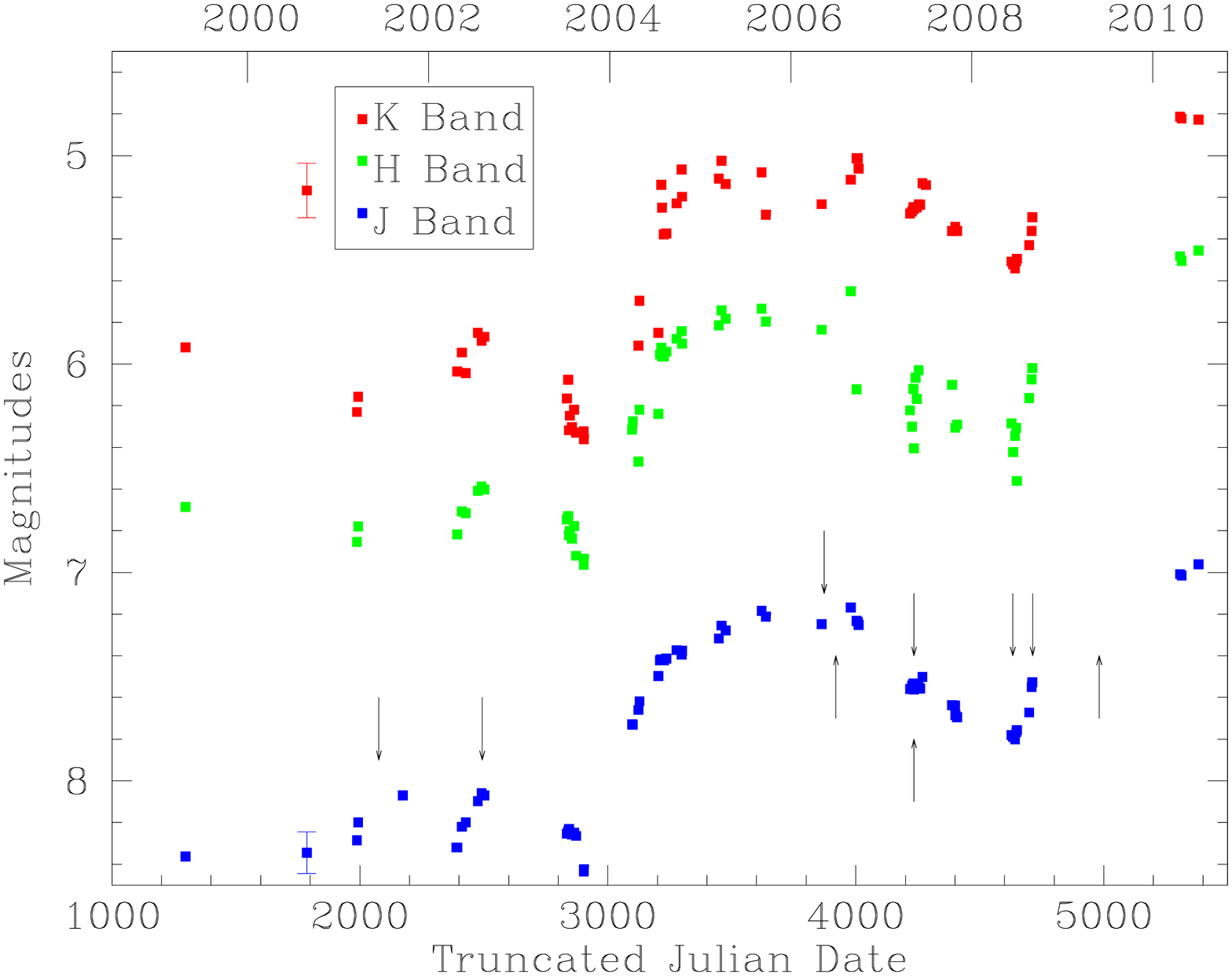}
\end{minipage}
\hfill
\begin{minipage}{8cm}
\centering
\includegraphics[width=6cm,angle=270]{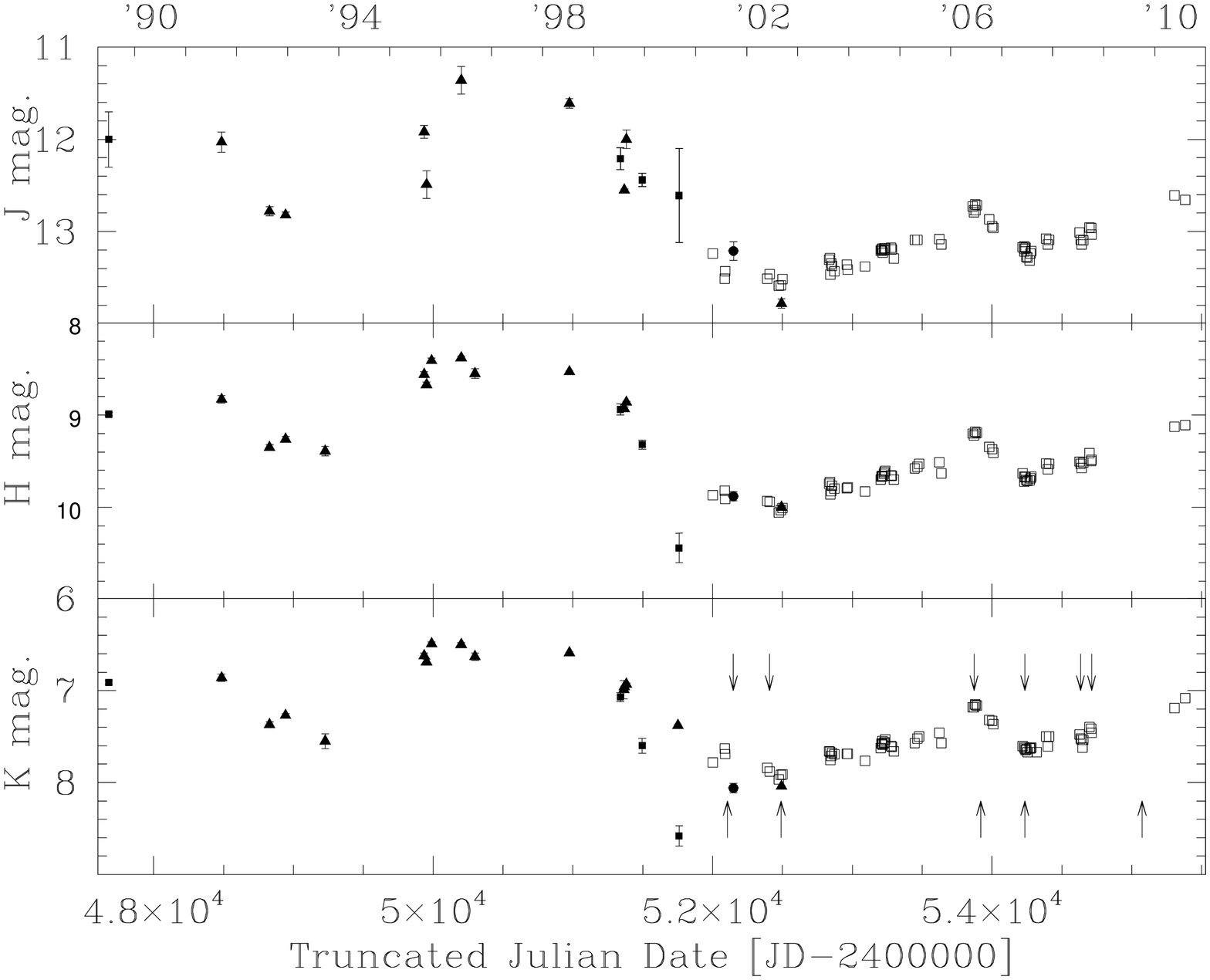}
\end{minipage}
\caption{ Long term JHK lightcurves of G24.73+0.69 (left panel; Clark et al. in prep) and AFGL2298 (right panel; Clark et al. 2009). Times
of spectral observations are indicated by arrows.}
\end{figure}

\begin{figure}[h]
\centering
\includegraphics[width=17cm]{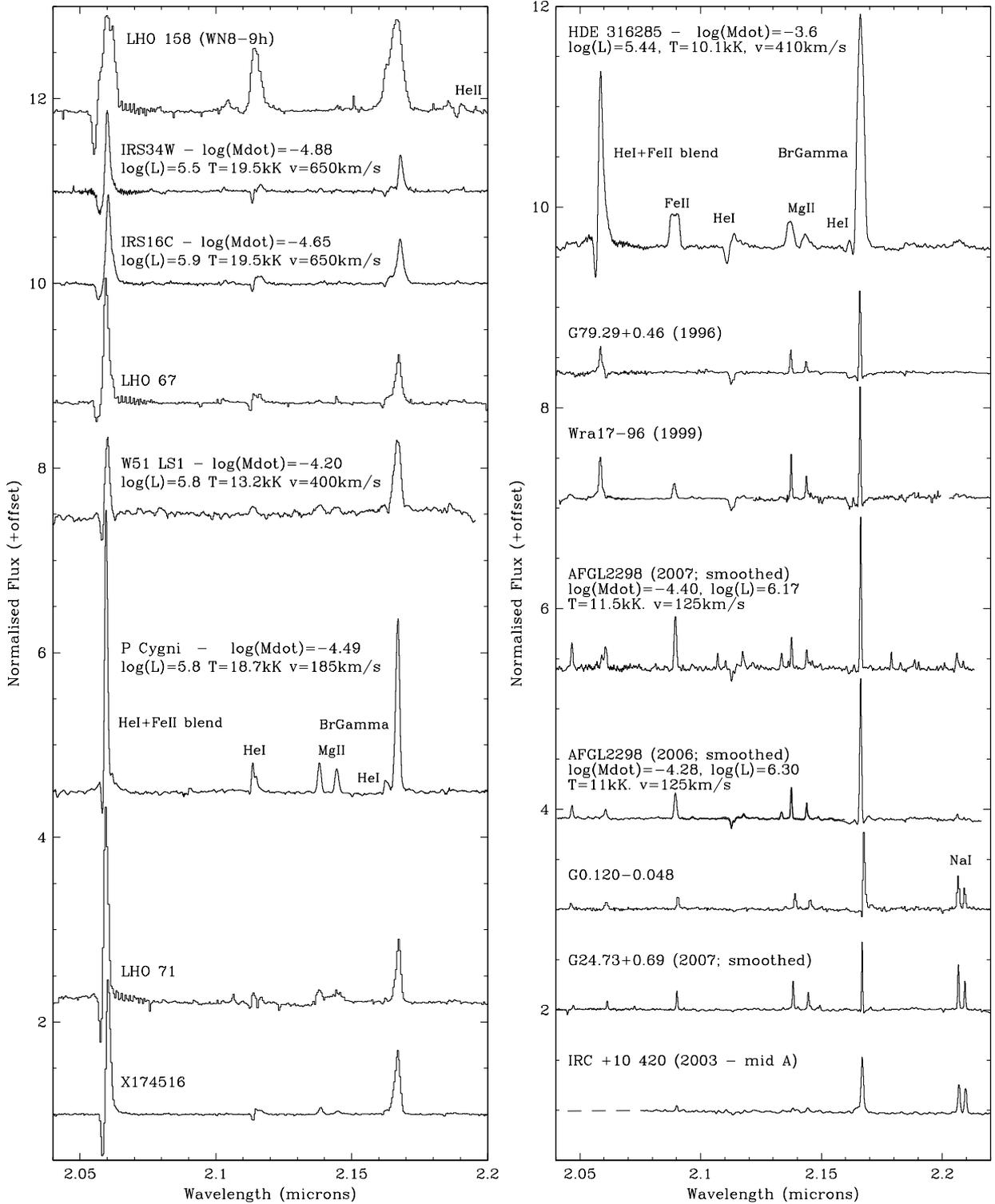}
\caption{Montage of K band spectra of galactic (candidate) LBVs demonstrating their diverse morphologies.
For comparison the spectra of the Yellow Hypergiant IRC +10 420 and the WN8 Quintuplet member LHO 158 are also presented.  
 For reasons of brevity the references to the origin of spectra and associated analyses have been omitted but are
given in Clark et al. (in prep.).  
}
\end{figure}

%
% USE A SECTION WITHOUT NUMBER FOR THE ACKNOWLEDGEMENTS
%
\section*{Acknowledgements}

AZT-24 observations  made within
an agreement between Pulkovo, Rome and Teramo observatories.
%
% BEGIN THE REFERENCE LIST WITH \beginrefer
% USE \refer BEFORE THE REFERENCES AND BEGIN A NEW PARAGRAPH AFTER THE 
% REFERENCE !
% DO NOT FORGET TO END THE LIST WITH \endrefer
%
\footnotesize
\beginrefer
\refer Clark, J. S., Egan, M. P., Crowther, P. A., et al., 2003, A\&A, 412, 185

\refer Clark, J. S., Larionov, V. M., Arkharov, A., 2005, A\&A, 435, 239

\refer Clark, J. S., Crowther, P. A., Larionov, V. M., et al., 2009, A\&A, 507, 1555

\refer Clark, J. S., Ritchie, B. W., Negueruela, I., 2010, A\&A, 514, 87

\refer Crowther, P. A., Smith, L. J., 1996, A\&A, 305, 541

\refer Crowther, P. A., Pasquali, A., DeMarco, O., et al., 1999, A\&A, 350, 1007

\refer Crowther, P. A., Hadfield, L., Clark, J. S., Negueruela, I., Vacca, W. D., 
2006, MNRAS, 372, 1407

\refer Gal-Yam, A.,Leonard, D. C., 2009, Nature, 458, 865 

\refer Groh, J. H., Hillier, D. J., Damineli, A., et al., 2009, ApJ, 698, 1698

\refer Gvaramadze, V. V., Kniazev, A. Y., Fabrika, et al., 2010, MNRAS, 405, 1047

\refer Hanson, M. M., Conti, P. S., Rieke, M. J., 1996, ApJS, 107, 281

\refer Hillier, D. J., Miller, D. L., 1998, ApJ, 496, 407.

\refer Hillier, D. J., Crowther, P. A., Najarro, F., Fullerton, A. W., 1998, A\&A, 340, 483

\refer Langer, N., Hamann, W.-R., Lennon, M., et al. 1994, A\&A, 290, 819 

\refer Liermann, A., Hamann W.-R., Oskinova, L. M., 2009, A\&A, 494, 1137

\refer Martins, F., Hillier, D. J., Hillier, D. J., et al. 2007, A\&A, 478, 219

\refer Mauerhan, J. C., Morris, M. R., Cotera, A., 2010, ApJ, 713, L33

\refer Najarro, F., Figer, D. F., Hillier, D. J., Geballe, T. R., Kudritzki, R. P., 
2009, ApJ, 691, 1816 

\refer Smith, N., Owocki, S., 2006, ApJ, 645, L45

\refer Wachter, S., Mauerhan, J. C., Van Dyk, S. D., et al., 2010, AJ, 139, 2330

\endrefer           
\end{document}